\documentclass[final,5p,times,twocolumn]{elsarticle}

\usepackage{amssymb}
\usepackage[T1]{fontenc}
\usepackage{xcolor}
\newcommand{\url}[1]{\texttt{#1}}

 \usepackage{lineno}

 \biboptions{comma,square,sort&compress}

\usepackage{amsmath}
\DeclareMathOperator{\ZZ}{ZZ}
\newcounter{bla}

\makeatletter
\renewcommand{\ps@pprintTitle}{}
\makeatother

\begin{document}

\begin{frontmatter}



\title{ZZPolyCalc: An open-source code with fragment caching for determination of Zhang-Zhang polynomials of carbon nanostructures}


\author[a]{Rafa{\l} Podeszwa\corref{author}}
\author[b]{Henryk A. Witek}
\author[b]{Chien-Pin Chou\fnref{chienpin}}

\cortext[author] {Corresponding author.\\\textit{E-mail address:} rafal.podeszwa@us.edu.pl}
\address[a]{Institute of Chemistry, University of Silesia in Katowice, Szkolna 9, 40-006 Katowice, Poland}
\address[b]{Department of Applied Chemistry and Institute of Molecular Science, National Yang Ming Chiao Tung University, Hsinchu 300093, Taiwan}
\fntext[chienpin]{Current address: Materials Informatics Initiative, RD Technology and Digital Transformation Center, JSR Corporation, 3-103-9, Tonomachi, Kawasaki-Ku, Kawasaki, Kanagawa, 210-0821, Japan}

\begin{abstract}
Determination of topological invariants of graphene flakes, nanotubes, and fullerenes constitutes a challenging task due to its time-intensive nature and exponential scaling. The invariants can be organized in a form of a combinatorial polynomial commonly known as the Zhang-Zhang (ZZ) polynomial or the Clar covering polynomial. We report here a computer program, ZZPolyCalc, specifically designed to compute ZZ polynomials of large carbon nanostructures. The curse of exponential scaling is avoided for a broad class of nanostructures by employing a sophisticated bookkeeping algorithm, in which each fragment appearing in the recursive decomposition is stored in the cache repository of molecular fragments indexed by a hash of the corresponding adjacency matrix. Although exponential scaling persists for the remaining nanostructures, the computational time is reduced by a few orders of magnitude owing to efficient use of hash-based fragment bookkeeping. The provided benchmark timings show that ZZPolyCalc allows for treating much larger carbon nanostructures than previously envisioned.

\end{abstract}


\end{frontmatter}


{\bf PROGRAM SUMMARY}

\begin{small}
\noindent
{\em Program Title: } ZZPolyCalc                                         \\
{\em Developer's repository link:} {\scriptsize \url{https://github.com/quantumint/zzpolycalc}} \\
{\em Licensing provisions:} GPLv3  \\
{\em Programming language:} Fortran 2008, C                                  \\
{\em Nature of problem:}
  Calculations of Zhang-Zhang (ZZ) polynomials for carbon nanostructures have traditionally been performed by the application of a recursive algorithm, which scales exponentially with the size of the system. For large nanostructures, the problem often becomes intractable due to prohibitive computational cost. Consequently, ZZ polynomials were reported so far only for small or model systems, for which analytical formulas could be determined; larger systems, such as graphene flakes, fullerenes, or nanotubes, have been intractable by the existing implementations. \\ 
{\em Solution method:} 
  The process of determination of ZZ polynomials, based on recursive decomposition of a given carbon nanostructure, produces many identical fragments. In ZZPolyCalc, we avoid redundant calculations for such fragments by storing all the intermediate data in a hash table and reusing it efficiently whenever necessary. This new algorithm is particularly effective for elongated graphene flakes or carbon nanotubes, where the scaling becomes polynomial, but it also greatly reduces the computational time for other, less regular nanostructures.  
\end{small}

\section{Introduction}
\label{s:intro}

The Zhang-Zhang (ZZ) polynomial approach \cite{Zhang:96a,Zhang:97} is a generalization of the chemical resonance theory \cite{Pauling:77} widely used in organic chemistry to predict stability and reactivity of aromatic compounds. The existing implementations \cite{Chou:12, Chou:14} of the ZZ polynomial  algorithm \cite{Zhang:96a,Gutman:06b} were useful to produce many interesting results \cite{Chou:14b,Witek:15a,Langner:18b,He:21d} for small and highly structured, medium-size benzenoids, but it seems that their limit of applicability has been reached with regard to larger carbon nanostructures such as graphene flakes \cite{Witek:21d}, fullerenes, and nanotubes.  

The current paper reports a Fortran 2008 code (ZZPolyCalc) designed for fast and automatic determination of ZZ polynomials, which are computed via recursive decomposition of the molecular graph of a given nanostructure. The usual top-to-bottom decomposition of the molecular graph \cite{Gutman:06b, Chou:12}, traditionally used to determine ZZ polynomials, has been augmented in the current implementation by a sophisticated bookkeeping algorithm, in which every fragment (subgraph) appearing in the recursive decomposition is stored in the cached repository of molecular fragments together with the partial ZZ polynomial corresponding to the analyzed fragment. The bookkeeping algorithm is based on unique  labels obtained by hashing of the adjacency matrix of the fragment subgraph. The presented benchmark results show that ZZPolyCalc allows for treating much larger systems than the previous implementations of the ZZ polynomial codes \cite{Chou:12, Chou:14}.

The overhauled implementation reported in the current work should allow one to apply the ZZ polynomial theory to study various properties of large benzenoids, including their energetic stability, equilibrium geometries, aromaticity, reactivity, spin-density maps for radicals, nucleus-independent chemical shift (NICS) indices \cite{Schleyer:96}, and all other properties for which the aromatic character has an impact. For large systems, for which the standard methods of quantum chemistry become too expensive computationally, one of the viable options is using topological and graph-theoretical invariants, which often provide qualitatively correct information about the studied systems \cite{Gutman:06c}.  Note that usual fragmentation techniques of quantum chemistry \cite{aoki:12,fedorov:12,li:12} might not be suitable here due to very long conjugation length associated with the network of double and aromatic $\pi$ bonds in large carbon nanostructures. However, our code based on the resonance theory reasoning
can be useful in guiding \cite{Noffke:20} the fragmentation schemes, in addition to providing chemically valuable information benefiting the traditional quantum chemistry based investigations.

\section{Theoretical background}
\label{s:theory}

\begin{figure}[t] 
  \begin{center} 
    \includegraphics[width=1.0\linewidth]{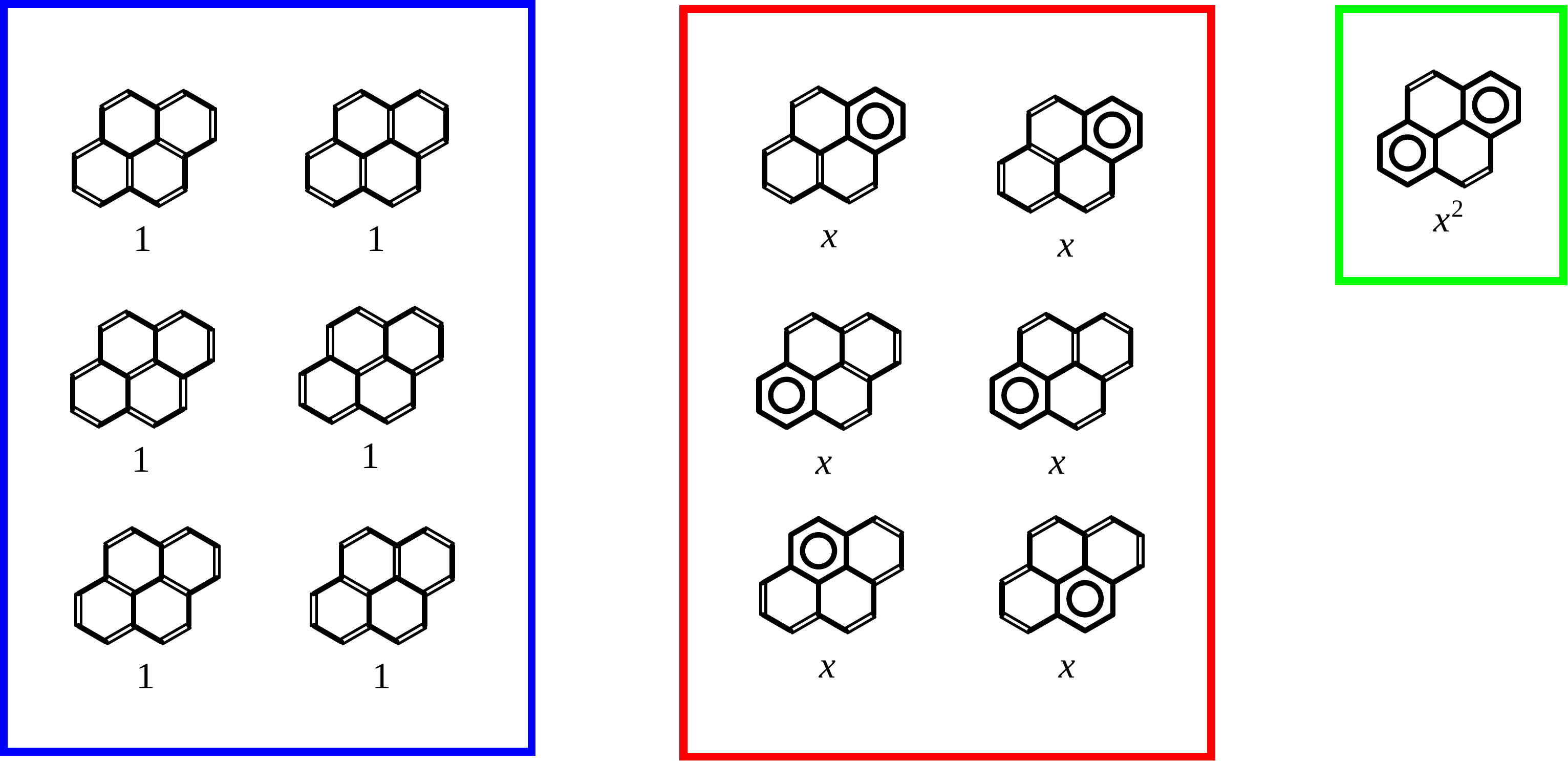} 
  \end{center}
  \caption{The complete set of Clar covers of pyrene. For details, see text.} 
  \label{fig:pyrene} 
\end{figure}

Zhang-Zhang polynomials ({\em aka} Clar covering polynomials or ZZ polynomials) offer a convenient way of enumerating Clar covers that can be constructed for a given benzenoid. In the language of resonance theory, a Clar cover  of a benzenoid $\mathbf{B}$ is a generalized resonance structure describing the $\pi$ bonding system of $\mathbf{B}$ in terms of double bonds \includegraphics[width=0.035\linewidth]{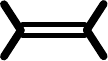} (abbreviated in the following as $K_2$) and Clar sextets  {\includegraphics[width=0.025\linewidth,angle=0]{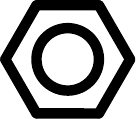}} (abbreviated in the following as $C_6$). Various arrangements of double bonds $K_2$ and aromatic sextets $C_6$ in $\mathbf{B}$ give rise to distinct Clar covers of $\mathbf{B}$.  The number of Clar sextets $C_6$ in a given Clar cover $C$, denoted as $\mathrm{ord}\left(C\right)$, is referred to as the order of $C$. The maximal order of all Clar covers
\begin{equation}
  \label{eq:Cl}
  Cl = \max \left\{\mathrm{ord}\left(C\right) | \; C\,\in\, \left\{\mathrm{Clar\;covers\;of}\;\mathbf{B}\right\} \right\}  
\end{equation}  
is referred to as the Clar number of $\mathbf{B}$. The Clar covers of order 0 correspond to the well-known Kekul\'{e} structures of $\mathbf{B}$. The most convenient way to familiarize the reader with the concept of Clar covers proceeds via building a complete set of Clar covers for some particular benzenoid. To this end, in Figure~\ref{fig:pyrene} we have constructed all the Clar covers of pyrene \includegraphics[width=0.028\linewidth]{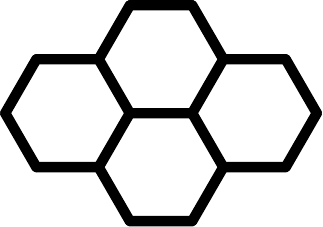}. 
The Clar covers are grouped by their order: the structures in the blue frame are purely Kekul\'{e}an, the ones with one aromatic ring are in the red frame, and the only member of the two-ring family is marked with a green frame. 

The Zhang-Zhang polynomial $ZZ\left(\mathbf{B} ,x\right)$ is a combinatorial polynomial enumerating  Clar covers of $\mathbf{B}$ of distinct orders. Formally, the definition of the ZZ polynomial is given by
\begin{equation}
  \label{eq:zz}
  ZZ\left(\mathbf{B} ,x\right)\hspace{8pt}= \hspace{-25pt} \sum_{C\,\in\, \{\mathrm{Clar\;covers\;of}\;\mathbf{B}\}}   \hspace{-30pt} x^{\,\mathrm{ord}\left(C\right)}
\end{equation}
Since every Clar cover of order $k$ gives the same contribution $x^k$ to the ZZ polynomial, it is often more convenient to express $ZZ\left(\mathbf{B} ,x\right)$ via an alternative formula equivalent to Eq.~(\ref{eq:zz}); we have
\begin{equation}
  \label{eq:ZZ}
  ZZ\left(\mathbf{B} ,x\right)=\sum_{k=0}^{Cl} c_k \,  x^k 
\end{equation}
where the sequence $[c_0, c_1, \ldots, c_{Cl}]$ lists the number of Clar covers in each order class. The contributions constituting the ZZ polynomials of pyrene are printed under each Clar cover shown in Figure \ref{fig:pyrene}. There are six possible Kekul\'{e} structures with order 0, six Clar covers with one aromatic sextet, and one Clar cover with two aromatic rings. Consequently, the Clar number of pyrene is 2, and its ZZ polynomial is $ZZ\left(\mathrm{pyrene} ,x\right)=6+6x+x^2$. 

The idea of representing the number of Clar covers as a polynomial may sound overly complicated since the sequence of coefficients $[c_0, c_1, \ldots, c_{Cl}]$ is equivalent to the ZZ polynomial. However, it turns out that it is easier to calculate the whole ZZ polynomial than a single coefficient $c_i$. This feature stems from the recurrence properties of the ZZ polynomials \cite{Zhang:96a,Gutman:06b,Chou:12} based on the decomposition of the molecular graph tree. The resulting recursive algorithm allows one to compute the ZZ polynomial of $\mathbf{B}$ as a sum of ZZ polynomials of subgraphs of $\mathbf{B}$. The computation time of the ZZ polynomials scales exponentially with the increasing size of the system. Moreover, for large systems,  the calculations tend to overflow the standard 4- or 8-byte integer limits and have to be performed with arbitrary arithmetic precision.

\section{Possible applications of the code}
\label{s:applic}

Zhang-Zhang polynomials were introduced to chemical graph theory over a quarter-century ago to generalize the concept of the number of Kekul\'{e} structures. In those days, it became more and more evident that relying only on the number of Kekul\'{e} structures of a given benzenoid $\mathbf{B}$ (usually referred to as the Kekul\'{e} count of $\mathbf{B}$ and abbreviated as $\mathcal{K}\equiv\mathcal{K}\left(\mathbf{B}\right)$) to predict the stability and reactivity of aromatic compounds might not be adequate. Therefore, various generalizations of Kekul\'{e} structures started to be studied (original Clar idea of conjugated sextets \cite{Clar:72}, Randi\'{c} conjugated circuits \cite{Randic:76}, Pleter\v{s}ek generalized Clar covers \cite{Pletersek:18}), which possibly better describe the stability of various $\pi$-bonded nanostructures in the spirit of Pauling's chemical resonance theory. These new concepts worked better than the mere determination of the Kekul\'{e} count $\mathcal{K}$ of a given benzenoid and relating it to the $\pi$ energy of this structure. For example, it was found that the most stable isomer of $\rm C_{60}$ buckminsterfullerene with $I_h$ symmetry does not maximize the Kekul\'{e} count $\mathcal{K}$ among for all the pentagon/hexagon-bearing isomers of $\rm C_{60}$ \cite{Austin:94}: There are 20 isomers with a larger value of $\mathcal{K}$ than the $I_h$ isomer with no apparent correlation between the thermodynamic stability and the Kekul\'{e} count $\mathcal{K}$. The difference in the Kekul\'{e} count $\mathcal{K}$ is substantial for the $I_h$ isomer, with the Kekul\'{e} count 24\% lower than for the isomer with the maximum Kekul\'{e} count. Zhang and coworkers later discovered that the stability of the $I_h$ isomer could be explained by taking into account not only the Kekul\'{e} count $\mathcal{K}$ but also the Clar number $Cl$ \cite{Zhang:10}. The $I_h$ isomer has the largest number of Clar covers among structures with the largest Clar number, $Cl=8$, with 10\% higher count than the second-highest such an isomer. The aromaticity of $I_h$ is also open to a heated debate, with some studies suggesting that this isomer is anti-aromatic \cite{Chen:12}. In the recent study of the $(5,6)$-fullerenes C$_n$ with $n = 20$--$50$  \cite{Witek:20a}, no simple rule relating thermodynamic stability and the Kekul\'{e}/Clar counts has been found, suggesting that such a relation is not going to be trivial and thus increasing the motivation for the present coding effort. Note that the topological information clearly influences the energetic stability of fullerenes \cite{Sure:17}: The most stable $I_h$ isomer is the only isomer of $\rm C_{60}$, which is in compliance with the isolated pentagon rule (IPR), which states that fullerenes with no adjacent pentagons are more stable than those with adjacent pentagons.  

Another interesting observation made in Ref.~\cite{Witek:20a} is related to the fact that all the 811  ZZ polynomials for fullerenes $\rm C_{20}$--$\rm C_{50}$ are unique; extending  this observation to larger fullerenes (or disproving such a relationship) should be possible with ZZPolyCalc. The ZZ polynomial, being invariant to the structure representation, constitutes a convenient unique label that could be used to differentiate between the plethora of fullerene isomers, whereas the actual form of the adjacency matrix depends on the atom numbering. This feature is critical in machine learning, as shown in an atomization energy study, where a unique molecular structure descriptor enhanced the accuracy of the prediction tremendously \cite{Hansen:13} compared to the previously used Cartesian coordinates based descriptors. 

The idea of predicting geometry using Kekul\'{e} structures goes back to Pauling bond orders, defined as an average of $\pi$ bond orders (0 or 1) over all the conceivable  Kekul\'{e} structures of a given benzenoid. For pyrene, shown in Fig.~\ref{fig:pyrene}, the resulting Pauling bond orders are $1/6$, $2/6$, $2/6$, $3/6$, $3/6$, and $5/6$, and the corresponding optimized DFT bond lengths \cite{Fadli:17} decrease almost monotonically (1.435~\AA, 1.431~\AA, 1.425~\AA, 1.408~\AA, 1.396~\AA, and 1.367~\AA, respectively), giving reasonable agreement with the bond-order predictions \cite{Herndon:74b}.  This effect is not coincidental, and the bond orders and the geometrical patterns agree well for many hydrocarbons \cite{Herndon:74a, Herndon:74b}. The described here protocol can be readily extended to Clar covers assumming that the bond order in a Clar sextet {\includegraphics[width=0.025\linewidth,angle=0]{benzenerot.pdf}} is $1/2$. For pyrene, the resulting Clar bond orders are $3/26$, $10/26$, $6/26$, $14/26$, $13/26$, and $23/26$, giving six different values with slightly worse agreement with the trends in the optimized DFT bond lengths than the Kekul\'{e} results. 
Clearly, the bond length predictions from the resonance-type methods are not perfect. Recently, Gutman and coworkers found several examples of hydrocarbons, where the Pauling bond order predictions turned out to be incorrect with the simplest example of the chevron molecule \cite{Radenkovic:14}, where the bond orders for the central bonds are monotonically decreasing. In contrast, the bond lengths calculated with DFT behave more irregularly, with a shallow minimum and a rapid increase of the bond length for the terminating bond at the bay-position in the chevron molecule. ZZPolyCalc could be useful to decide whether Clar bond orders can be useful in resolving such discrepancies. ZZPolyCalc could be also used to generate qualitatively correct optimal geometries of fullerenes. The current method of generating 3D geometries of fullerenes from their topology is based on the spiral algorithm, which produces the corresponding adjacency matrix \cite{Sure:17}. To obtain the geometry of a given isomer, one needs to project the adjacency matrix on the surface of a 2D manifold, which in many cases leads to very strained molecular geometries originating from the requirements to place a certain number of atoms in a very confined region. Using such a high-energy strained molecular geometry as an input for a quantum chemical optimization program may result in a computational failure. Generation of molecular geometries via ZZPolyCalc possibly coupled with machine learning algorithms would likely avoid such problems.

Another application of Clar covers is to describe aromaticity.  Consider the example of pyrene shown in Fig.~\ref{fig:pyrene}. Pyrene contains two pairs of equivalent hexagons: outer and inner. The outer hexagons have aromatic character in three Clar covers and the inner hexagons have aromatic character in only one Clar cover, which gives raise to three times larger aromaticity in the outer ring than in the inner ring. The corresponding ratios, $3/13$ and $1/13$, can be considered as the Clar measures of aromaticity. These values can be compared to the archetypal aromatic system, benzene, for which the analogous ratio is $1/3$. Therefore, the outer rings of pyrene have strong local aromaticity, close to that of benzene, whereas the inner rings are only weakly aromatic. Using benzene's aromaticity as a reference, we can define percentage Clar aromaticity of the outer and inner hexagon, which are 69\% and 23\%, respectively. This observation agrees with the local aromaticities determined using the harmonic oscillator model of aromaticity (HOMA) \cite{Krygowski:00}: 0.86 for the outer rings and 0.70 for the inner rings \cite{Ghosh:11}, and similar predictions obtained with the NICS aromaticity index: $-11.28$ and $-3.92$, respectively \cite{Ghosh:11}. (Strongly negative value of NICS corresponds to a pronounced aromatic character of a ring.)  One can obtain the same conclusions for many polyaromatic hydrocarbons. For example, in coronene, there are only 2 (out of 69) Clar covers with the aromatic sextet at the central ring, for which NICS predicts strongly non-aromatic value of $-0.7$ \cite{Ciesielski:06}, consistent with the current Clar predictions. The aromatic character of the central ring alternates between the high and low values with growing size of the flakes. This observation is fully supported by the Clar analysis \cite{Page:03}, where the same patterns with similar quantitative changes are predicted. 

It is important to mention that the ZZ polynomials have also applications in mathematics. The ZZ polynomials are equivalent to the so-called cube polynomials \cite{Zhang:13, Berli:15, Pletersek:18}, and are related to the strict order polynomials of partially ordered sets (posets) \cite{Langner:21a,Langner:22a,Langner:22b,Langner:22c} sextet polynomials \cite{Zhang:00}, and certain tiling polynomials \cite{Langner:17,Langner:18}. An interesting line of research concerns discovering closed-form formulas for structured families of benzenoids, which would allow one to study various asymptotic properties of these structures \cite{Witek:21}. Two approaches has been pursued there, one based on a solution to a set of recurrence relations derived for a number of structurally related benzenoids generated by the recursive decomposition of a given family \cite{Chou:12c,Chou:14b,Chou:14c,Chou:15,Witek:15a,Chou:16,Witek:17,Langner:18b,He:20,He:21b}, and the second based on heuristic enumeration of Clar covers for subsequent members of a given family in order to discover a formula describing such set of ZZ polynomials \cite{Chou:12b,He:21}. ZZPolyCalc can be particularly useful for generating new results in the latter approach by allowing one to perform a series of ZZ polynomial calculations for various families of open-ended and capped nanotubes, large fullerenes, and elongated graphene flakes with isostructural edges. The information deduced from numerical analysis of a series of ZZ polynomials of such structures with increasing size could be particularly useful for discovering closed-form ZZ polynomial formulas for various classes of carbon nanotubes, i.e.,  systems that could not be analyzed before due to the prohibitively long computational time.

\section{Technical details}
\label{s:technical}

A thorough explanation of the recursive decomposition algorithm utilized for the calculation of ZZ polynomials was provided in Ref.~\cite{Chou:12}. Here, we find it instructive to exemplify this algorithm by depicting in Figure~\ref{fig:anthanthrene} two initial steps in the decomposition process of anthanthrane (\textbf{B1} in Figure~\ref{fig:anthanthrene}).  Multiple alternative pathways can be explored as well; however, in the instance at hand, the algorithm sequentially addresses bonds from left to right, designating the bond selected for the decomposition with a dot. The designated bond can be assigned a single (S), double (D), or aromatic ring (R) character. The bond nature imposes certain bonding restrictions on either the entire molecule or its fragment. These constraints are depicted in gray in Figure~\ref{fig:anthanthrene}. The initial decomposition step of \textbf{B1} results in fragments \textbf{B2}, \textbf{B3}, and \textbf{B4}; all these fragments are different. In the subsequent step, ZZPolyCalc decomposes \textbf{B2} producing fragments \textbf{B5}, \textbf{B4}, and \textbf{B6}, for which the ZZ polynomials are computed in the subsequent steps (not shown in Figure~\ref{fig:anthanthrene}) and saved in the cached repository of fragments. Subsequently, ZZPolyCalc decomposes \textbf{B3} producing fragment \textbf{B4} and two completely decomposed structures. There is no need to compute the ZZ polynomial for any of these fragments: The ZZ polynomial for the decomposed structures are 1 and $x$, and the ZZ polynomial for the fragment \textbf{B4} can be recovered from the repository. Similar observation concerns the last step, in which the ZZ polynomial of \textbf{B4} is again read from the repository.
The resulting ZZ polynomial of \textbf{B1} can be eventually computed from the obtained data as follows:

\begin{figure*}[t]
\begin{center}
\includegraphics[width=0.9\linewidth]{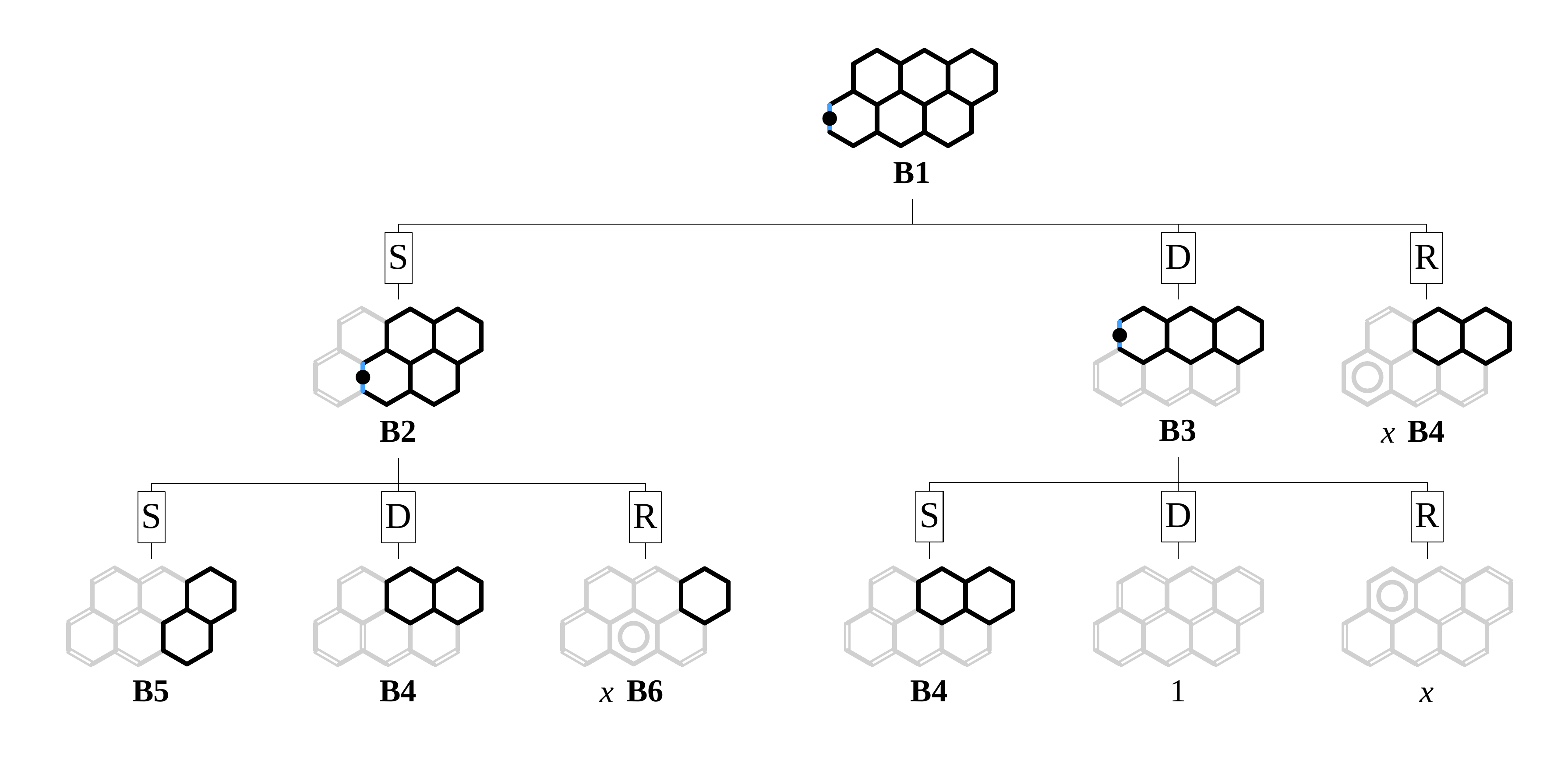}
\end{center}
\vspace{-20pt}
\caption{Two initial steps in the recursive decomposition of anthanthrene show that fragment \textbf{B4} appears multiple number of times in the decomposition tree.}
\label{fig:anthanthrene}
\end{figure*}

\begin{equation*}
\ZZ(\textbf{B1},x)=\ZZ(\textbf{B2},x)+\ZZ(\textbf{B3},x)+x \ZZ(\textbf{B4},x) 
\label{eq:B1}
\end{equation*}
where the ZZ polynomials
\begin{align*}
\ZZ(\textbf{B2},x)=&\ZZ(\textbf{B5},x)+\ZZ(\textbf{B4},x)+x \ZZ(\textbf{B6},x) \\
\ZZ(\textbf{B3},x)=&\ZZ(\textbf{B4},x)+1+x
\end{align*}
and $\ZZ(\textbf{B4},x)$ are all read from the cached repository. Note that fragments \textbf{B5} and \textbf{B4} are identical and share the same polynomial, a fact that we have decided not to exploit in ZZPolyCalc: Numerical tests showed that identification of isomorphic graphs and/or design of a hashing algorithm invariant to permutation of graph vertices \cite{Portegys:15} had too high computational complexity to be useful in the process of speeding up the ZZ polynomial calculations. Consequently, the fragments \textbf{B5} and \textbf{B4} are not considered identical by ZZPolyCalc and both are stored in the cache library as unique fragments.

Designing an efficient algorithm for locating fragments in the repository can be challenging. A simple and fairly efficient way to address this problem is by assigning a uniquely determined label (hash) to each fragment; such hash would point to the ZZ polynomial of a given fragment in a dynamically-updated hash-indexed repository of ZZ polynomials. The hashes are fairly small and can be computed quickly for arbitrarily large data. In ZZPolyCalc, one can choose either popular cryptographic hash algorithms from Secure Sockets Layer (SSL) libraries (MD5 \cite{Rivest:92} or SHA-256 \cite{Barker:08}), or very fast XXHASH128 \cite{Collet:19}. Too short hashes increase the risk of collisions, but with 128-bit hashes (MD5 or XXHASH128), the likelihood of two structures having the same hash is astronomically small \cite{Bellare:05}, even if one considers billions of entries of cached data (in our largest calculation so far with $3 \times 10^{9}$ of fragments, the collision probability is about $1.3\times 10^{-20}$).  Therefore, even with significantly larger systems than the current version of ZZPolyCalc can treat, there is no practical need for 256-bit hashes (SHA-256). Although neither MD5 nor XXHASH128 are cryptographically secure, the quasi-random distribution of hashes in our implementation of ZZPolyCalc should not lead to any discernible problems with collisions.

\begin{figure}[t]
\begin{center}
\includegraphics[width=0.6\linewidth]{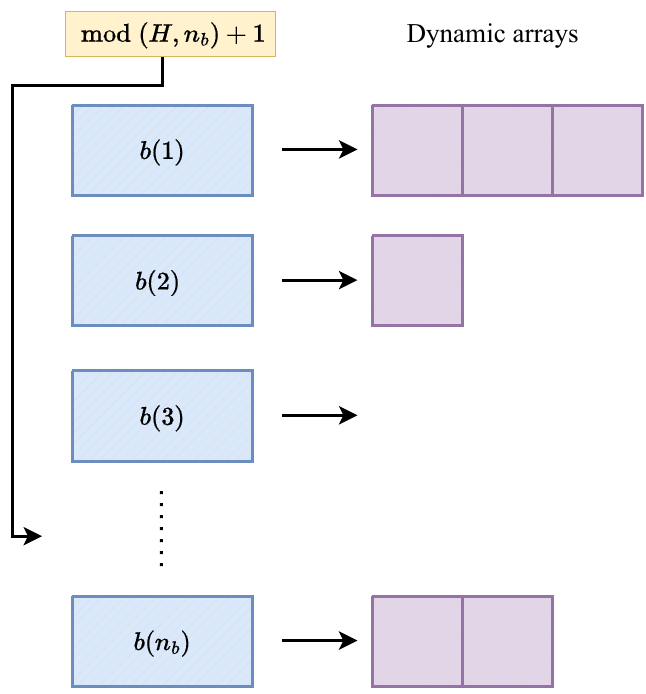}
\end{center}
\caption{The organizational structure of the cache database with $n_b$ buckets used in ZZPolyCalc. Each bucket contains a pointer to a dynamic array, that can vary in size (including zero). A fragment with hash $H$ resides in bucket $b\left(\bmod (H, n_b)+1\right)$.}
\label{fig:cache}
\end{figure}

During the recursive ZZ polynomial calculations, the hash of each fragment is computed and compared against the hashes already stored in the dynamic repository, organized as a variation of a hashed array tree with $n_b$ buckets $b(i), i=1\ldots n_b$. Each fragment, uniquely identified by a hash $H$ computed from its adjacency matrix, is stored in the bucket $b\left(\bmod (H, n_b)+1\right)$. Every bucket maintains a pointer to a dynamic array of its content, which expands as new fragments are appended to a given bucket. For a sufficiently large $n_b$, most buckets contain not too many fragments, that minimizes the resource-intensive operations on large dynamic arrays. The structure of the cache database is depicted in Figure~\ref{fig:cache}.

For larger systems, the sheer number of fragments may exhaust the available memory. In such instances, we purge the repository of fragments that are deemed less important. The significance of a fragment $f$ is assessed via a score function
\begin{equation}
    S\left(f\right)=\frac{\text{age}\left(f\right)}{\text{freq}\left(f\right)\,\left[\text{size}\left(f\right)\right]^2}
\end{equation}
where $\text{age}\left(f\right)$ is the count of fragments added to the repository after the most recent read of $f$, $\text{size}\left(f\right)$ is the number of atoms in $f$,  and $\text{freq}\left(f\right)$ is the number of times the fragment $f$ has been used (counting the initial use as one). Once the number of stored fragments reaches the database limiting size, the fragment with the highest score is replaced by a newer one. This method penalizes fragments that are small, infrequently used, or old. We chose to perform the purge within the same bucket to which the new fragment is assigned, maintaining the simplicity of the database structure. Provided that the average number of structures per bucket is substantial—--a condition induced by choosing $n_b$ not too large--—there is a sufficient selection of structures eligible for replacement. This approach is particularly effective for quasi-1D moieties, like carbon nanotubes or elongated hexagonal graphene flakes. Using this method, we managed to compute the ZZ polynomial for a $10 \times 10 \times 1000$ flake with $40\,200$ atoms while maintaining a database of only 12 million unique fragments out of more than 3 billion processed fragments. A more sophisticated algorithm for fragment removal could reduce memory requirements even further, as a significant portion of structures is never reused and could thus be discarded without any consequence.

ZZPolyCalc offers the user an opportunity of saving the database cache to disk at the end of the computation or at regular intervals. The latter option is useful for checkpointing large calculations in order to restart the code later, e.g., after a power outage. The saved cache can also be utilized for a different  member of the same family of structures, as the fragment repositories of both members contain many identical fragments allowing one for an efficient fragment reuse.

The efficiency of the ZZ polynomial recursive algorithm, as well as the size of the resultant fragment set, depends on the traversal path chosen during the computation. We observed that for a quasi-1D nanostructure oriented along the $z$ direction, the most effective strategy involves processing atoms in the order determined by successive sorting of their respective  Cartesian coordinates in the order $z,y,x$ (set as a default in ZZPolyCalc). Following this observation, the two alternative optimal orientations of the graphene flake shown in Figure~\ref{fig:structures}, which  minimizes the computational time are as follows. $(i)$ The flake lies in the $zy$ plane with the $z$ axis parallel to the arrow $n$. $(ii)$ The flake lies in the $yx$ plane with the $y$ axis parallel to the arrow $n$. Choosing an appropriate order of atoms in the nanostructure can have enormous effect on the calculation process: An optimal ordering can reduce the computational time by a few orders of magnitude. For example, the computational time for the flake $6\times 6 \times 20$ in orientation $(ii)$ sorted in the order $z,x,y$ is 175 times longer than for the analogous flake sorted in the order $z,y,x$.   Alternative decomposition pathways might be pursued with ZZPolyCalc by providing the user-supplied atom order list, which depreciates the default sorting and determines the order of subsequent decompositions in the traversal path. 

\section{Numerical results}

All timings reported in this study were obtained using an Intel Xeon CPU E5-2660 v3 running at 2.6~GHz. The ZZPolyCalc program was compiled with the Intel Fortran Compiler version 19.0.117, and the xxHash library version 0.8.2 was utilized. Unless otherwise specified, the number of buckets $n_b$ was set to $2\,097\,152$, and the maximum number of records in the database was capped at $52\,000\,000$. 

\begin{figure}[t]
\begin{center}
\includegraphics[width=0.40\linewidth]{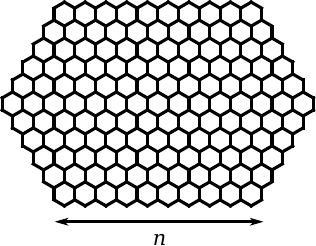}\hspace*{2em}
\makebox[0.4\linewidth][l]{\hspace*{0cm}\raisebox{0.05\linewidth}{\includegraphics[width=0.40\linewidth]{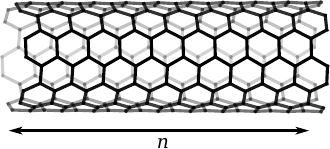}}}
\end{center}
\caption{Visualization of the $6\times 6 \times n$ hexagonal graphene flakes and the $(6,6)$-$n$ armchair nanotubes utilized in this study.}
\label{fig:structures}
\end{figure}

\begin{figure}[t]
\begin{center}
\includegraphics[width=1.0\linewidth]{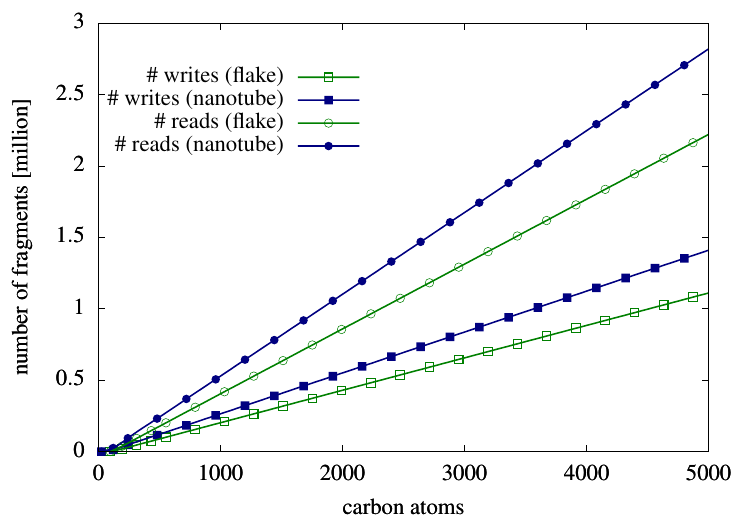}
\end{center}
\caption{Number of fragments processed during recursive decomposition of the $(6,6)$-$n$ carbon nanotubes (solid symbols) and the $6\times 6\times n$ hexagonal graphene flakes (open symbols). The label `\#~writes' represents the number of unique fragments saved to the cache database, while `\#~reads' denotes the number of fragments restored from the cache database during the ZZ polynomial computation process. The number of carbon atoms in the studied nanotubes and flakes is $24n$ and $24n+72$, respectively.}
\label{fig:structperatom}
\end{figure}

Figure~\ref{fig:structperatom} illustrates the number of cached fragments for a series of $6\times 6 \times n$ hexagonal flakes and $(6,6)$-$n$ carbon nanotubes. (Molecular models of these nanostructures are shown in Figure~\ref{fig:structures}.) The number of database operations---and consequently: the size of the cached fragment repository---grows linearly with the size of the analyzed nanostructures, except for very small systems. This very feature is the main advantage of ZZPolyCalc over the previous ZZ 
implementations, as it allows us to reduce the overall scaling of the calculations from exponential to polynomial for quasi-1D systems. Note that for $n>6$, the number of new fragments added to the repository, when transitioning from calculations for a nanostructure of length $n$ to an analogous nanostructure of length $n+1$, is constant and equals to $6870$ for the $(6,6)$-$n$ carbon nanotubes and $5448$ for the $6\times 6 \times n$ hexagonal flakes. Interestingly, the number of operations in which a new fragment is added to the repository constitutes $1/3$ of all database operations, despite the fact that some of the fragments are accessed hundreds of times and other fragments (over 60\% of them), never. We hypothesize that the ratio $1/3$ is closely related to the branching ratio of the molecular graph in the decomposition process [assignment of a single (S), double (D), or aromatic ring (R) character to the bond at the decomposition point] as discussed earlier in Section~\ref{s:technical}.

\begin{figure}[t]
\begin{center}
\includegraphics[width=1.0\linewidth]{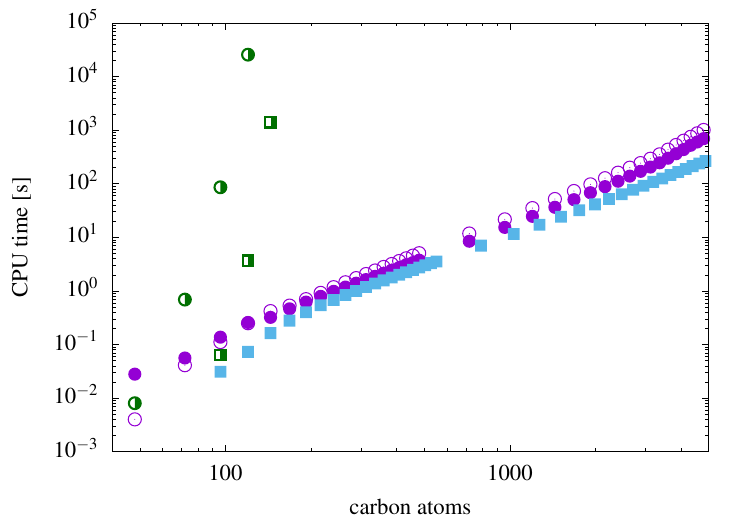}
\end{center}
\caption{CPU time for computing the ZZ polynomials of the $(6,6)$-$n$ carbon nanotubes (circles) and the $6\times 6 \times n$ hexagonal graphene flakes (squares). The half-filled symbols, corresponding to ZZCalculator (the previous implementation of the non-cached code from Ref.~\cite{Chou:12}), suggest exponential scaling. In contrast, ZZPolyCalc (filled and open symbols) time scales polynomially with the number of carbon atoms. Filled circles 
indicate the use of a large number of buckets ($n_b=2\,097\,152$) and open circles to small number of buckets ($n_b=1000$); for details, see text.}
\label{fig:nttiming}
\end{figure}

In Figure~\ref{fig:nttiming}, we present a comparison between the timing performance of ZZPolyCalc and ZZCalculator (the previous implementation of the non-cached code from Ref.~\cite{Chou:12}) for structures shown in Figure~\ref{fig:structures}. Despite the minor overhead associated with establishing the internal database of cached fragments (a situation that can be mitigated for small systems by reducing the number of buckets in the database, see Figure~\ref{fig:nttiming}), ZZPolyCalc scales polynomially for quasi-1D systems. Since both the number of cached fragments and their average size (impacting, e.g., hashing time) grow linearly with the increasing size of the system, the overall scaling of ZZPolyCalc must be at least quadratic. However, additional factors---associated with the length of the ZZ polynomial representation---lead to a further increase in scaling, especially for the $(6,6)$-$n$ carbon nanotubes, which is evident by the increased slope of the timings in Figure~\ref{fig:nttiming}. The ZZ polynomial computation complexity is slightly higher for the nanotubes than for the  $6\times 6 \times n$ hexagonal graphene flakes. The reason for the difference is twofold. First, for $n>10$, the ZZ polynomial order of each flake is 36 regardless of its length. However, there is no analogous limit for the nanotubes. For example, the ZZ polynomial order for a $4800$-atom nanotube with $n=200$ is as large as 798. Secondly, the number of digits in the ZZ polynomial coefficients for the flakes grows modestly with $n$, achieving 68 for $n=200$, whereas the analogous value for the nanotube is 412.

It is important to stress that ZZCalculator (the non-cached code from Ref.~\cite{Chou:12}) becomes impractical for larger systems; we estimate that for the $(6,6)$-$6$ nanotube with 144 atoms, it would take several months to complete the computation with ZZCalculator. In contrast, ZZPolyCalc performs the same task within a fraction of a second. It is also worth noting that a reduced number of buckets in the database, while advantageous for small systems, results in increased computational cost for larger ones. In such a situation, the dynamic array associated with each bucket becomes sizeable and all the operations on it become somewhat slower compared to the rapid selection of an appropriate bucket in scenarios with multiple buckets.

\begin{figure}[t]
\begin{center}
\includegraphics[width=1.0\linewidth]{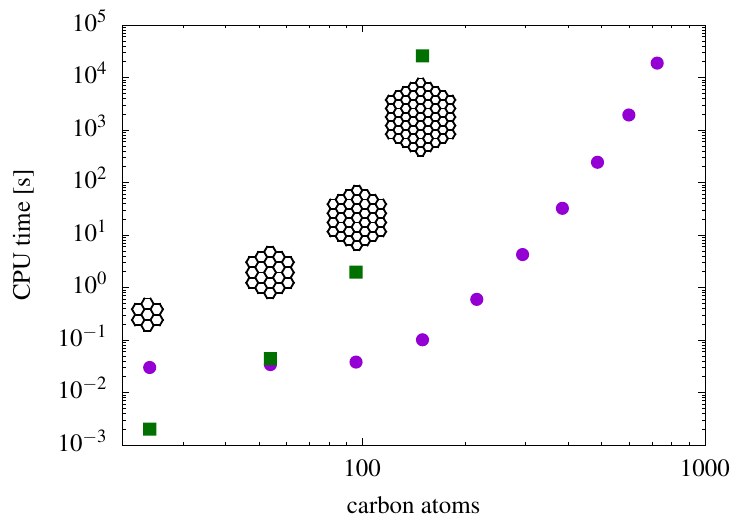}
\end{center}
\caption{CPU time for computing the ZZ polynomials for a series of circumcoronenes. The circles correspond to the ZZPolyCalc code from this work, while the squares refer to the older ZZCalculator code from Ref.~\cite{Chou:12}. Schematic circumcoronenes models are depicted for the first four members of the family.}
\label{fig:circumcoronenes}
\end{figure}

For non-quasi-linear systems, such as fullerenes or circumcoronenes, the scaling of the ZZPolyCalc remains exponential. Nevertheless, the fragment caching strategy brings significant benefits. This is well illustrated in Figure~\ref{fig:circumcoronenes}, where we compare the performance of ZZPolyCalc and ZZCalculator for a family of circumcoronenes. Using ZZCalculator, the computation time for circumcircumcircumcoronene (C$_{150}$H$_{30}$) exceeded 4 hours; we have estimated that for the subsequent member of this family (C$_{216}$H$_{36}$), we would need several years to complete the calculations. In contrast, the cached ZZPolyCalc code allows for the same computation to be completed in a fraction of a second, while the computation for the largest considered here member of this family, C$_{726}$H$_{66}$, takes approximately 5 hours.

\section*{Declaration of Competing Interest}

The authors declare that they have no known competing financial interests or personal relationships that could have appeared to influence the work reported in this paper.

\section*{Acknowledgements}

RP acknowledges COST Action CA21101 “Confined molecular systems: from a new generation of materials to the stars” (COSY) supported by COST (European Cooperation in Science and Technology). HAW and RP acknowledge support from the Taiwan Ministry of Science
and Technology (MOST 110-2923-M-009-004-MY3 and  NSTC111-2113-M-A49-017).

\section*{Data availability}

The molecular geometries, calculated ZZ polynomials, and the timings for the systems used in the numerical results presented here are available in the \texttt{examples} directory of the source code.

\bibliographystyle{elsarticle-num}
\bibliography{iii}







\end{document}